# Accessible Adventures: Teaching Accessibility to High School Students Through Games

**Kyrie Zhixuan Zhou** — University of Illinois Urbana-Champaign, United States | zz78@illinois.edu

**Chunyu Liu** — University of Illinois Urbana-Champaign, United States | chunyu3@illinois.edu

**Jingwen Shan** — University of Illinois Urbana-Champaign, United States | jshan10@illinois.edu

**Devorah Kletenik** — Brooklyn College, City University of New York, United States | kletenik@sci.brooklyn.cuny.edu

**Rachel F. Adler** — University of Illinois Urbana-Champaign, United States | radler@illinois.edu

## ABSTRACT
Accessibility education has been rarely incorporated into the high school curricula. This is a missed opportunity to equip next-generation software designers and decision-makers with knowledge, awareness, and empathy regarding accessibility and disabilities. We taught accessibility to students (N=93) in a midwestern high school through empathy-driven games and interviewed three Computer Science high school teachers and one librarian who taught programming. Accessibility education is currently insufficient in high school, facing challenges such as teachers' knowledge and conflicted curriculum goals. The students exhibited increased knowledge and awareness of accessibility and empathy for people with disabilities after playing the games. With this education outreach, we aim to provide insights into teaching next-generation software designers about accessibility by leveraging games.

## INTRODUCTION

Information and communications technologies (ICTs) are often designed without explicitly considering accessibility, effectively creating barriers to use for people with disabilities. Accessibility education is important to equip future software designers and developers with awareness and knowledge of accessibility and empathy for disabilities. It has been widely covered in both Computer Science (CS) curricula (Carter & Fourney, 2007; Wald, 2008) and non-CS curricula (Kurniawan et al., 2010) in college. In particular, game-based approaches have been effective in teaching accessibility. Kletenik and Adler developed games to simulate different types of disabilities, the empathy-inducing and educational effects of which were demonstrated through testing with undergraduate CS majors and non-CS majors (Kletenik & Adler, 2022, 2023, 2024).

Compared to the rich literature on accessibility education in college, only a handful of literature discusses accessibility education for K-12 students. Kelly and El-Glaly taught high schoolers about a web content accessibility guideline, Reflow, through a hands-on online module (Kelly & El-Glaly, 2021). Adler and Kletenik introduced accessibility games in a university-level web development course taken by K-12 teachers who taught or planned to teach CS courses (Adler & Kletenik, 2023). The teachers thought the games would benefit elementary, middle, and high school students. In contrast to accessibility education, the literature on AI ethics and cybersecurity education for this younger audience is rich, emphasizing topics taught (Ali et al., 2019), methods used (Forsyth et al., 2021), and tools and games developed (Chapman et al., 2014).

There are two reasons why incorporating accessibility into K-12 education is important. Firstly, students tend to consider accessibility as an afterthought instead of tightly integrating it within the design and development cycle (Edwards et al., 2006; Patricia, 2011; Conn et al., 2020). This is compounded by the fact that students' first exposure to accessibility usually comes many years after they have learned to program and after they have become accustomed to inaccessible technology design. Accessibility education in higher education thus requires undoing already-established bad software development habits. Secondly, research suggests that no single intervention results in long-lasting changes in student attitudes towards accessibility (Zhao et al., 2020; Conn et al., 2020). Rather, accessibility needs to be included throughout the educational trajectory, as early and as frequently as possible so that it is viewed not as a special topic but as a natural part of software development. For both reasons, we see the inclusion of accessibility in K-12 CS education as the necessary next step in training software designers and developers in accessible design.

To equip high school students with knowledge, awareness, and empathy about software accessibility, we used games in the accessibility education literature (Kletenik & Adler, 2022, 2023, 2024) to teach accessibility in CS and non-CS classes in a midwestern public high school. By delivering game-based, empathy-driven accessibility education to high school students, and interviewing CS teachers, we aspired to bridge the research gap in educating pre-college students about accessibility in an engaging way and probe the possibility of incorporating accessibility education into the high school curriculum. Specifically, we answered the following two research questions (RQs):

- How do students' empathy, knowledge, and awareness regarding accessibility change after playing the games?

- How is and how should software accessibility be taught in high school?

Our contributions are two-fold. First, we expanded engaged scholarship in accessible computing to high school education and innovatively used games to teach accessibility to high school students. We show the preliminary potential of equipping next-generation software designers and developers with knowledge, awareness, and empathy regarding accessibility as early as high school. Second, we discussed with teachers how to best integrate computer accessibility into the current high school curriculum.

# RELATED WORK

## Accessibility Education in Universities

In 2016, Putnam et al. only found one stand-alone course about accessibility (Putnam et al., 2016). In recent years, however, researchers have devoted extensive efforts to understanding objectives and principles of accessibility education in higher education, especially computing education (Conn, 2019; Kang et al., 2021), developing new pedagogical methods such as games (Kletenik & Adler, 2022, 2023, 2024), and understanding how students perceive accessibility education (Chávez & Van Wart, 2023). Researchers and industrial practitioners such as Baker et al. (2023) advocate accessible computing education in universities, framing accessibility as a cultural competence in computing. A systematic literature review found awareness of accessibility (e.g., abilities, laws, ethics), technical knowledge (e.g., requirements, guidelines, WCAG, testing), empathy (e.g., understating disabilities, inclusive design), and potential endeavors (e.g., pursuing career in accessibility) were four common learning objectives; HCI, Web Design, and Web Programming courses were more likely to include accessibility education; and the most frequently used pedagogies were in-class activities, projects, and lectures (Baker et al., 2020). Gellenbeck (2005) argued that similar to ethics education, the integration of accessibility education into the CS curriculum should follow four recommendations: (1) early introduction; (2) continued discussion in most courses; (3) integration of topics within the courses; and (4) maximum coverage with minimum overlap.

Accessibility is taught in a wide range of scenarios in universities. A comprehensive survey (N=1,857) with computing and information science faculty found that 175 institutions in the U.S. had at least one instructor teaching accessibility (Shinohara et al., 2018). Focused courses such as Accessible Computing have been taught to both CS students (Carter & Fourney, 2007; Wald, 2008) and non-CS students (Kurniawan et al., 2010). Accessibility has also been widely integrated into courses such as HCI (Liffick, 2004) and Web Design (Rosmaita, 2006). Siegfried & Leune (2022) shared a dedicated full-semester Accessibility Seminar. Besides teaching universal design (UD) and accessibility in CS-related courses and programs, some departments also provide relevant project work, summer courses, and master thesis topics (Nishchyk & Chen, 2018). Notably, faculty who were female, had expertise in HCI and software engineering, or knew people with disabilities were more likely to teach accessibility in computing and information disciplines (Shinohara et al., 2018).

Empirical evidence showed the success of integrating accessibility education into higher education computing curriculum; specifically, gains in awareness and knowledge regarding accessibility occurred when accessibility lectures were part of the course (Ludi et al., 2018). Chávez & Van Wart (2023) unpacked how students in a web development course perceived accessibility. Students responded to two open-ended questions embedded in a lab and homework: "Why, and to whom, is accessibility important?" "Do you think that designing for accessibility also improves the usability of the site for all users? Why or why not?" 62.5% of the students perceived accessibility as beneficial for all users of a website. However, a survey (N=114) with final-year computing undergraduates revealed that they did not personally view accessibility training as essential career preparation (Conn et al., 2020), inviting more research to study the long-term effect of accessibility education.

Various teaching methods have been used to teach accessibility, with in-class activities, projects, and lectures being the most common (Baker et al., 2020). For example, technical projects such as "implementing a Braille translator capable of converting English sentences into Grade 2 Braille representation using dictionaries" were integrated into core CS courses (Kuang et al., 2024): these assignments successfully increased students' familiarity with accessibility concepts yet failed to cultivate a mindset of accessibility. In addition to traditional pedagogical methods, experiential labs and simulation games have been developed to build empathy in students and facilitate experiential learning regarding accessibility and disabilities. El-Glaly et al. (2020) developed a set of Accessibility Learning Labs using an experiential learning structure to simulate the experience of people with disabilities and teach students to repair accessibility issues using a simulated code editor. These labs have been incorporated into computing-related courses (Shi et al., 2023). Gamification was proposed as a potentially engaging way to teach accessibility (Lorgat et al., 2022). Simulation games suitable for beginner CS students and non-majors without a coding component but with non-jargon accessible design guidelines have also been developed (Kletenik & Adler, 2022, 2023, 2024). After playing these simulation games, students showed increased empathy for disabilities and intention to design accessibly; they also had more accessibility ideas, such as including people with disabilities when testing design (Kletenik & Adler, 2024).

**Accessibility Education outside Universities**

Literature has been abundant on accessibility education in higher education, discussing learning objectives (Conn, 2019), short-term (Ludi et al., 2018) and long-term (Conn et al., 2020) outcomes, traditional teaching methods (Baker et al., 2020), and game-based teaching (Kletenik & Adler, 2022, 2023, 2024). However, since accessibility education has long been confined to universities, the technology sector's accessibility skills gap is persisting, highlighting "the need for academia and the workplace to learn from each other and adapt together to generate pedagogies that will better prepare learners for accessibility practice" (Coverdale et al., 2022).

One approach to enhancing practitioners' inclusive thinking and accessibility knowledge is to cultivate these values early. Kletenik and Adler argued that "accessibility should be taught as early as possible to CS-majors and non-majors" to enable them "to build awareness of the need to design accessibly and to increase empathy for people with disabilities" (Kletenik & Adler, 2022). K-12 teachers were receptive to including accessibility topics in their classrooms (Adler & Kletenik, 2023). However, few studies have focused on teaching accessibility in high school education or earlier. Kelly & El-Glaly (2021) designed an online module combining an interactive lecture and quizzes to demonstrate Reflow, one of the web content accessibility guidelines, to high school students, successfully increasing students' awareness of disabilities and knowledge of accessibility. Accessibility education through games (Kletenik & Adler, 2023) or gamification (Lorgat et al., 2022) is naturally more approachable and engaging to K-12 students, which has not been investigated to our knowledge.

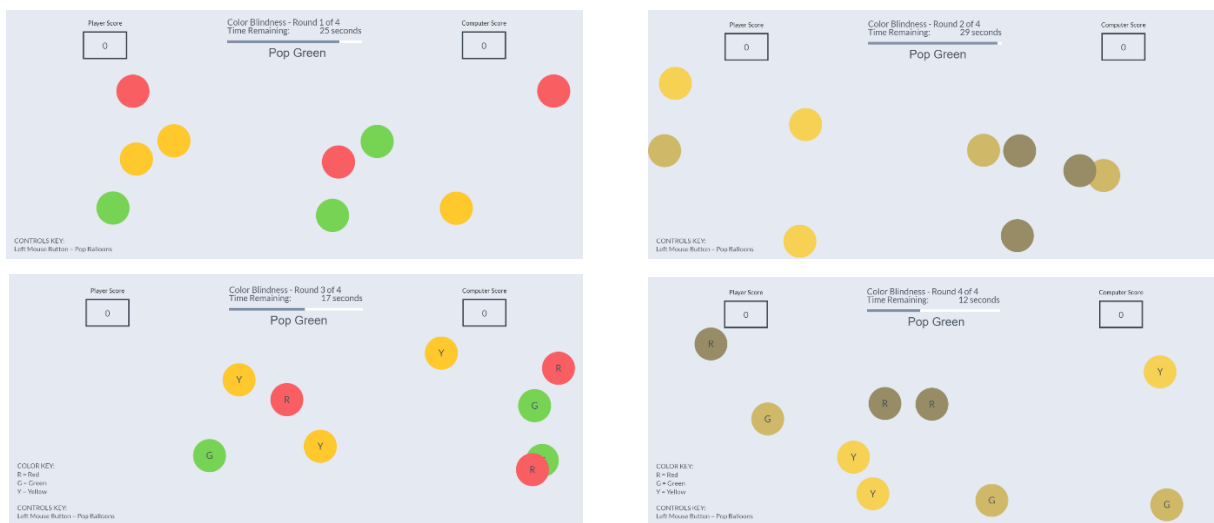

**Figure 1. Four rounds of the Color Blindness Game (Kletenik & Rachel, 2022).**

We bridge the research gap through the current study, teaching high school students about accessibility through simulation games developed by Kletenik & Adler (2022, 2023, 2024). These simulation games cover a wide range of disabilities, including Color Blindness, Auditory Impairment, Physical Impairment, Blindness, Low Vision, and Dyslexia. Each game has four rounds. In the first round, users are instructed to select and pop balls of specific colors (or words in the Dyslexia Game) without a disability condition; In the second round, users play the game with conditions that simulate the software challenges faced by people with disabilities (i.e., inability to tell colors in the Color Blindness Game, unable to hear colors to pop in the Auditory Impairment Game, having a shaky mouse in the Physical Impairment Game, unable to see balls in the Blindness Game, playing with blurry vision in the Low Vision Game, and seeing words with the phonemes remapped on the balls in the Dyslexia Game); In the third round, users play the game without a disability but with enhanced accessibility features (i.e., labeling balls with letters representing respective colors in the Color Blindness Game, additional text instructions in the Auditory Impairment Game, an additional keyboard input method in the Physical Impairment Game, additional audio instructions in the Blindness Game, a zooming bar in the Low Vision Game, and accompanying a word with a corresponding picture in the Dyslexia Game); In the fourth round, users play the game with conditions faced by people with a disability and with enhanced accessibility features. Four rounds of the Color Blindness Game are demonstrated in Figure 1 as an example. Guidelines to make software more accessible are summarized in the end. In the context of color blindness, for example, guidelines include: Don't use color alone for identification; Use sufficient contrast to make the text stand out from the background; Underline links and don't rely only on the color to indicate a link; Use descriptive text or alternate text to indicate color if it is important (for example, in color choices for clothing).

## METHODOLOGY

### Teaching Activities

We taught students about accessibility in a selective enrollment midwestern public high school including students from 8th graders (pre-freshmen) to 12th graders, and interviewed Computer Science teachers about their experience and perception of teaching accessibility to students.

Three CS courses included Computer Literacy 1, Computer Literacy 2, and Digital Media Productions. Computer Literacy 1 is a mandatory introduction course for 8th graders to learn concepts such as computer use, media literacy, AI, library service, HTML/CSS, and Python. Computer Literacy 2 is a mandatory course for freshmen to implement projects such as games and drones in teams. Digital Media Production is an elective course. We also utilized Coding Club (a lunchtime student activity), a Study Hall session, and two non-CS courses, i.e., Theater Class and Stage Craft.

The teacher informed students and their parents about our IRB-approved teaching activity, which allowed students and/or their parents to opt out. In the end, a total of 93 students participated in our activities and completed our study. Demographic information of the participating students is shown in Table 1.

| Grade | | Gender | | Race | | Class | |
|---|---|---|---|---|---|---|---|
| 8th Grade | 35 | Female | 39 | American Indian or Alaska Native | 4 | Coding Club | 3 |
| 9th Grade | 25 | Male | 48 | Asian | 25 | Computer Literacy 1 | 35 |
| 10th Grade | 14 | Other | 4 | Black or African American | 5 | Computer Literacy 2 | 26 |
| 11th Grade | 9 | Prefer not to say | 2 | Hispanic or Latino | 7 | Digital Media Productions | 7 |
| 12th Grade | 10 | | | Prefer not to say | 1 | Stage Craft | 6 |
| | | | | Two or more races | 16 | Study Hall | 8 |
| | | | | White | 35 | Theater Class | 8 |

Table 1. Demographic information of participating high school students (N=93).

*Student Game Playing and Classroom Discussion*

In each class, we first introduced ourselves and our intention to understand how students learn about digital accessibility. They were told about the voluntary nature of this activity and that we would not share anything that they wrote or said with their teachers, principals, or the school. We then introduced them to the six simulation games developed by Kletenik & Adler (2022, 2023, 2024). Students were prompted to play three of the six games.

Upon consent, students filled in a survey modified from Kletenik & Adler (2022) that asked about demographic information, CS/programming knowledge/experience, accessibility considerations when developing software, and 5-

point Likert-style questions. The Likert-scale questions included four attitude questions (a-d) and two challenge questions (e-f) (Kletenik & Adler, 2022). Following Likert-scale questions, we asked an open-ended question about design features to include when designing apps for people with disabilities (2). We also asked two open-ended questions about populations to consider when designing games (3) and voting booths (4), modified from Kletenik & Adler (2022). See these questions in Figure 2.

---

1. Please rate your level of agreement with the following statements: (5-point Likert scale – Strongly Disagree to Strongly Agree)
    a. Many current software applications are difficult for people with disabilities to use.
    b. People with disabilities are interested in new technology.
    c. A person with disabilities should not have to rely on someone around who can help.
    d. Software developers should provide technology suitable for use by people with disabilities.
    e. People with disabilities are likely to face challenges when interacting with many applications.
    f. If I design applications, I will try to keep in mind people with disabilities.
2. For example: If I design applications, I will try ...
3. Suppose you were coding a game and wanted to make sure that everyone could play it. Which sorts of people should you ask to try it out?
4. Suppose you were creating a voting booth for people to use in an election, and wanted to be sure that everyone could use it. Which sorts of people should you ask to try it out?

---

**Figure 2. Survey given before and after the games.**

After finishing the three games they chose, students were directed to the exit survey, which asked the same questions as in the pre-study survey, with extra open-ended questions eliciting feedback and suggestions to improve the games. The game-playing process took roughly 25 minutes in each class, and almost all students could finish three games and two surveys within this timeframe.

In the last ten minutes of each class except Coding Club and Theater Class, which had shorter time due to event conflicts, we facilitated a classroom discussion with several question prompts regarding prior teaching/learning of accessibility, current thoughts on accessibility challenges in software, learned strategies to improve digital accessibility, and perceptions of our teaching activity and the simulation games. The study session conducted with freshmen in Computer Literacy 2 had the most heated discussion and the richest student reactions while they played the games. The Computer Literacy 1 class with 8th graders was the quietest during game playing, possibly because the students were the youngest.

Two researchers took notes of student reactions and discussions during game playing and students' responses to the discussion questions at the end. Another researcher was designated to lead the discussions. No recordings were obtained to protect student privacy.

**Teacher Interviews**
We interviewed two full-time CS teachers, one part-time CS teacher, and a librarian who taught programming (three male and one female teachers) in this high school. The 45-minute interviews aimed to understand the courses and grades the CS teachers teach, if and how they covered computing and social/ethical concepts in their classes, if and how accessibility education was integrated into the CS curriculum, barriers to teaching accessibility, and teachers' own knowledge level in accessibility. The interviews were recorded and later transcribed for analysis.

**Data Analysis**
Through data triangulation, i.e., using different sources of data; method triangulation, i.e., using different methods of data collection or analysis; and investigator triangulation, i.e., using different researchers, we managed to corroborate findings and compensate for weaknesses in the data, thus increasing the validity and reliability of the results.

*Quantitative Analysis*
We employed statistical methods to understand how the disability modes affected students' performance and emotions and how the games improved student empathy and awareness regarding accessibility. We also compared our results to those in Kletenik & Adler (2022) to see if there were differences in educational effects between high school students and college students.

We analyzed students' emotions and performance in different rounds of the games. For each game round, we calculated the percentage of winning players and those who selected positive emotions ("relaxing," "fun," "enjoyable," "educational" as opposed to "boring," "frustrating," "difficult," and "confusing") in the question after playing each round within the game. For each accessibility game, we conducted a Cochran's Q test to see if statistical significance existed among rounds regarding winning and positive emotion percentages. For the statistically significant tests, we conducted post-hoc pairwise McNemar tests to identify significant round comparisons, employing the Bonferroni p-adjustment. We specifically examined whether Round 2 (disability mode) led to degraded performance and negative emotions and whether Round 4 (accessibility accommodations) improved performance and emotions.

To assess student empathy and intention to design accessibly, we used a two-tailed Wilcoxon signed-rank test with continuity correction to compare Likert-scale question responses in pre- and post-surveys.

For open-ended responses to the questions regarding population considerations for game design, population considerations for voting booth design, and feature considerations for app design, we first followed a qualitative content analysis approach (Krippendorff, 2018) to make them suitable for quantitative analysis. Two researchers coded the responses into either 0 or 1, where 0 indicated no awareness of accessibility, and 1 indicated some awareness of accessibility. For the game and voting booth questions, if students mentioned specific or general disabilities or impairments, we rated the responses as 1; for the app design question, if students mentioned specific design features such as larger font, we rated the responses as 1. We calculated Krippendorff's Alpha to assess inter-coder reliability between two independent coders and reached Krippendorff's Alpha of 0.792, 1.000, 0.922, 1.000, 0.925, and 0.978 for the pre-game, pre-booth, pre-design, post-game, post-booth, and post-design questions, respectively, averaging 0.936. This indicates a high degree of agreement between the two coders (Krippendorff, 2018). McNemar analysis was then employed to compare the pre- and post-survey responses with continuity correction.

To see if there were differences in students' accessibility perceptions between recreational (game) and duty-bound (voting booth) design, we used McNemar tests to compare responses to the game and voting booth questions. The analysis was carried out for both the pre-game survey and the post-game survey.

*Qualitative Analysis*
We adopted a thematic analysis for the qualitative analysis (Braun & Clarke, 2012). Two researchers independently conducted open coding of the open-ended survey responses, in-game comments about emotions, observation notes taken during the classes, and teacher interview transcripts. We regularly discussed to reach a consensus on emerging themes. For example, in analyzing teacher interviews, emerging themes included (1) social/ethical topics covered in CS classes, including cybersecurity, AI's social impact, and information literacy; and (2) accessibility education in high school, including learning goals, challenges encountered, and teachers' knowledge of accessibility. XMind, a mind-mapping tool, was used to organize different levels of themes and quotes into a hierarchical structure. Anonymized quotes will illustrate our findings. We did not calculate inter-coder reliability following best practices: because of the study's exploratory nature and the grounded coding process, which aimed to use coding as a process rather than as a product, we were interested in models and association, not quantification (McDonald et al., 2019).

**FINDINGS**
According to the open-ended response in the pre-survey which asked if students had created coding projects and what they looked like, 58 out of 93 students who participated in our study had learned programming language(s) or created apps/websites. This confirmed the importance and relevance of teaching software accessibility to high school students.

**Students' Perspectives**

*Game Performance and Emotions*
Cochran's Q test revealed a significant difference in the proportions of winning players across four rounds in the Auditory (Q(3) = 57.7, p <.0001), Color blindness (Q(3) = 37.6, p <.0001), Visual (Q(3) = 88.8, p <.0001), Low vision (Q(3) = 76.6, p <.0001), and Dyslexia (Q(3) = 96.1, p <.0001) games. Further, Pairwise McNemar tests showed that players faced significantly more difficulty playing Round 2 of these games, with lower proportions of winning players in this round than all other rounds (p <.05). No significant difference is revealed in different rounds of the physical game (Q(3) = 4, p = 0.261), echoing results in Kletenik & Adler (2022), since the disability mode with a mouse is harder, and more frustrating, but not impossible. Accessibility features, if implemented appropriately, helped students perform better in disability modes. Notably, students' performance in Round 4 was comparable to Round 1 in the Auditory, Color Blindness, Physical, and Dyslexia games. During game-playing, some students expressed that they did not like Round 4 of the Blindness game, which was still challenging for them without being able to see anything. In the classroom discussion, one student commented, "Still impossible even in Round 4 of the blindness game."

The rounds simulating disability made students anxious and frustrated. During game playing, the students were frustrated by the challenges faced by people with disabilities. When playing Round 1 (normal mode) of the games, they verbally commented, "Easy game." "See, five [points]." When playing Round 2 (disability mode) of the games, they were surprised by the level of difficulty people with disabilities had to face when using software, "It's so hard. What is that?" "Why is it like that?" "Oh dear, that's..." "It's frustrating!" They also reacted to different disability modes, e.g., "(Low vision) I'm tired." "(Physical) Shaky mouse? What? I'm so shaking." "(Blindness) I can't see anything [sigh]." "(Color blindness) I got negative scores! Click yellow? It's a bit challenging."

Survey analysis revealed significant differences among rounds in terms of emotions for all six games ($Q(3) = 36.0$, $p < .0001$ for Auditory, $Q(3) = 60.0$, $p < .0001$ for Color Blindness, $Q(3) = 25.6$, $p < .0001$ for Physical, $Q(3) = 57.8$, $p < .0001$ for Visual, $Q(3) = 59.3$, $p < .0001$ for Low Vision, and $Q(3) = 107.0$, $p < .0001$ for Dyslexia). For all six games, the percentage of positive emotions in Round 2 is significantly lower than in other rounds, as indicated by Pairwise McNemar tests ($p < .05$). Students' open-ended comments during game-playing explained these negative emotions: "(Low Vision) It was impossible to see what color I was supposed to click." "(Auditory) was literally impossible." "(Color Blindness) I felt my anxiety rise." "(Dyslexia) It was very stressful and confusing. It gave me a headache." "(Visual) I was just clicking randomly. It felt like I had no control over anything." "(Physical) I did not realize how hard it would be to play with this disability!" Except for the Blindness game, where Round 4 showed significantly fewer positive emotions than Round 1 ($p < .01$), and the Auditory game, where Round 4 showed significantly more positive emotions than Round 1 ($p < .05$), emotions in Round 4 of all other games were comparable to Round 1, without significant difference.

*Student Empathy and Awareness of Design Accessibility*
We explored the potential of disability simulation games to cultivate high school students' empathy for disabilities and their awareness and knowledge of accessibility. Overall, after playing games, the students showed increased knowledge, awareness, and empathy for disability and accessible design.

We observed statistically significant differences between the pre-game and post-game Likert-scale question responses: pre-mean and median of the attitude question responses were 3.91 and 3.75, respectively; post-mean and median increased to 4.20 and 4.25, respectively ($p < .0001$, effect size = .65). Similarly, pre-mean and median of the challenge question responses were 4.07 and 4.00, respectively; post-mean and median increased to 4.41 and 4.50, respectively ($p < .0001$, effect size = .59). Based on Cohen's classification of effect size, where 0.1 to 0.3 represents a small effect, 0.3 to 0.5 indicates a medium effect, and more than 0.5 suggests a large effect (Cohen, 1988), we observed statistically significant changes coupled with a large effect size regarding attitude questions and challenge questions according to Cohen's d. These findings suggest a substantive improvement in high school students' empathy for disabilities and their increased intention to design with a focus on accommodating the needs of people with disabilities, echoing Kletenik & Adler (2022).

After coding the game, voting booth, and app design responses into 0 or 1, indicating whether students explicitly considered people with disabilities or expressed concrete accessibility features, we compared pre- and post-responses utilizing McNemar analysis. In the game question, responses mentioning people with disabilities increased from 35 (37.6%) in the pre-survey to 83 (89.2%) in the post-survey ($\chi2 (1, N = 93) = 41.49$, $p < .0001$, effect size = .46). Similarly, in the voting booth question, responses including people with disabilities rose from 48 (51.6%) to 77 (82.8%) ($\chi2 (1, N = 93) = 23.76$, $p < .0001$, effect size = .44). In the app design question, there was an increase from 25 (26.9%) to 39 (41.9%) ($\chi2 (1, N = 93) = 5.63$, $p < .05$, effect size = .23) in responses mentioning people with disabilities. According to Cohen's g, a large effect size was observed in the game question and the voting booth question, whereas a medium effect size was observed in the app design question. Before playing the games, many students would only ask their family, friends, or classmates to try out designs they created. After game-playing, they were more inclined to include people with various disabilities, such as blindness and dyslexia, to test the designs. A few more students could indicate specific design features to accommodate accessibility challenges, which were taught in the games we used, e.g., "Make sure to put the name of the colors over them so people with colorblindness can see them."

During the classroom discussion, several students expressed their awareness regarding accessible design after playing the games, e.g., "I never thought about it [accessibility]. I often just overlook it. Now I think it's necessary to take it into account. It's not only for people with disabilities. It benefits everyone." "Previously I didn't understand what the captions were for on the websites. I thought it was not necessary. After playing the games, I find it important to make them more accessible for people with disabilities."

They also learned practical strategies to make software more accessible. Some of them shared their learned strategies to make software more accessible. For instance, they understood developers should ask people with disabilities for feedback/testing, "They [Developers] can consult people who have those disabilities"; and should play the simulation games to understand accessibility challenges, "The developers should play these games so they know the challenges". They also learned design features that could enhance accessibility, including different options for input/output, "Nice to have other options in addition to mice"; visual assistance for dyslexia, "Shouldn't only rely on words for dyslexia"; zooming affordances, "Zooming can be implemented easily but helps"; and personalized interface, "Personalization is really important. Dark mode is one. Mode to disable animations. Turn on the audio. A lot of games have a color-blind mode."

When asked how they would have designed the simulation games, the students emphasized the simulation of disability experience, e.g., "Present difficulty brought by different games and different disabilities"; connection to daily software use to inspire empathy, "Connecting it with common things is interesting, like navigating a website. People don't realize how difficult it is to process information. Helpful for empathy"; and integrating perspectives from people with disabilities, "Include personal angles from people with disabilities." Several students indicated they would take a similar empathy-driven approach to ours, "Understand who you're trying to educate. Go about similar ways you guys are demonstrating the difficulty faced by people with disabilities."

### Accessibility Perception in Different Design Scenarios

Kletenik & Adler (2022) found that students were more inclined to consider users with disabilities during the design of voting booths than games since they may subconsciously perceive individuals with disabilities as more actively engaged in voting activities than game participation. Similarly, we found that 51.6% of participants included people with disabilities in the voting booth question in the pre-survey, compared to 37.6% in the game design question – the difference is significant ($\chi^2$ (1, N = 93) = 7.58, p <.01, effect size = .34) with a large effect size based on Cohen's g. After engaging with the accessibility games, this distinction vanished: 82.8% of the students indicated accessibility considerations in the voting question, and 89.2% in the game question (p = 0.332).

### Feedback and Suggestions on Games

We analyzed open-ended responses at the end of the post-game survey to gauge the students' feedback and suggestions regarding our education activities. Most students gave positive feedback about the games, describing them as thought-provoking, fun, educational, informational, and easy to digest. One particular advantage of the games was that they could simulate software challenges for people with disabilities and let students see what software use is like for people with disabilities. This way, students were able to learn about disability challenges when software is inaccessible through the games, "It was an educational experience to help me see how hard it is for people with disabilities to use websites"; and learn how to make technologies more accessible, "I enjoyed learning how accommodating to people with disabilities makes technology more accessible and enjoyable." In terms of suggestions, some students wished the games were made more entertaining and complex, e.g., having a variety of games instead of being limited to the ball-popping game. The robot voice was annoying to some.

During the classroom discussion, the students similarly had an overall good experience with the games. They agreed that the instructions were understandable, nothing was confusing, and thought the games would "work well in K-12 contexts." They "liked the way it shows different types of disabilities." Some students expressed interest in playing the remaining three games other than the three they had played. They also provided suggestions for improving the games. For example, several students mentioned a bug in Round 4 of the Blindness game and other games that used audio features, where two colors are announced simultaneously in audio. According to other students, having audio for instructions, having more visuals such as pictures for the arrow keys, demonstrating how to play the game in the first round, and having games in other languages could make the games more understandable and accessible for younger kids in different countries.

### A Lack of Accessibility Education in High School

According to most of the students in the classroom discussion, accessibility education was not included in the current CS curriculum in high school. They shook their heads when asked if they had previously learned or thought about accessibility issues in computer software. Only one student mentioned that a brief discussion about accessibility occurred in a tech class in her middle school, and another student mentioned the teaching of accessibility principles in a university-level Psychology class. As a result, a lack of awareness of accessible designs was commonly expressed, "I tended to underestimate how difficult it would be." "I never thought about putting text in the color-blindness game." One student nevertheless implemented accessibility features in practice when she posted on social media, "I write a lot for the school's social media account. Different levels of visual impairments make social media difficult to use in different ways. I'd always add captions to images."

**Teachers' Perspectives**

*Social and Ethical Topics Covered*
CS courses covered several social and ethical topics, such as cybersecurity, AI's social impact, copyright, and information literacy. All four teachers covered cybersecurity in their classes. For example, T3 taught about responsible computing in daily use, given the prevalence of data breaches. T4 taught about malware and phishing broadly, not limited to financial risks, but also about safety and sexual risks. She tended to teach these concepts in a sociotechnical manner, often without coding. Two teachers taught about copyright. For instance, T3 covered a wider range of topics regarding copyright, including responsible use of content, fair use, licenses for code, and authorship, which, in his eyes, are "a part of the CS culture." Information literacy was a large part of T4's classes. In Computer Literacy 1, she taught about library and information, navigation, and information credibility assessment. AI's impact on society was taught by T1. He used videos from the Internet to teach different versions of AI, its history, its impact on life, and more recent advancements like ChatGPT. All teachers thought social and ethical issues were important topics to cover in CS classes as a part of "socially appropriate training" (T1). T4 was the co-teacher of CL1 and CL2 along with T1 and T3 and estimated a ratio of 60% vs. 40% regarding the time devoted to technical vs. social aspects in these two compulsory courses. Notably, the teachers did not mention accessibility when inquired about ethical and social topics covered.

*Accessibility Education*
According to the teachers, accessibility education was mostly covered in Web Dev courses and briefly mentioned across other CS courses, and topics included visual accessibility (T1, T2), auditory accessibility (T2), color blindness (T1, T2, T3), and multi-platform accessibility (T1). T2 tried to ensure students knew the basics of how to do HTML properly including making websites accessible in the Web Dev course. Similarly, T1 included a unit about website accessibility in his Web Dev course, teaching about colors and contrast, and accessibility generator and checker. He implemented hands-on practices for students to fix inaccessible areas on websites. According to him, students enjoyed building accessible websites, "Kids feel it cool that they are building unusual websites with different colors, not black and white." However, he acknowledged not discussing much about accessibility in other courses. Not including accessibility into compulsory courses such as CL1 and CL2, which emphasized other ethical and social topics, is a missed opportunity.

Teachers were divided regarding perceived barriers to teaching accessibility in CS courses. T1 thought there were not any barriers since no fancy materials in addition to websites and teacher notes were needed. T2 also did not think access to teaching materials was a barrier in his school but pointed out that some schools did not even have a computer lab. Different from these two optimistic teachers, T4 noted that each course usually lasts for one semester, and there is limited time to cover every possible topic in CS. She would choose to use a few standalone sessions instead of a separate unit to provide students with an overall awareness of accessibility. Alternatively, she might consider integrating accessibility throughout the curriculum rather than teaching it separately. Barriers expressed by T3 included the availability of tools for demonstration, teacher familiarity with the topic, and pedagogical engagement, i.e., how to present concepts to students in an engaging manner. He thought the games we used and our teaching activities could well address these barriers, particularly for their directness, "Students are not being told what to do, but directly experiencing the fact."

Among the four teachers, only T2 perceived himself as having adequate knowledge about accessibility due to his education and teaching experience. He had taken tech or education courses in college, which explicitly taught about accessibility and diversity. He also had a mindset of keeping learning about accessibility, "I understand it, but there's always more to understand. Continual learning and adjusting is important." Other teachers had not received formal education about accessibility and were less confident about their knowledge level in this regard. T3 learned accessibility from coding and life experiences. For example, he had friends and family members who had color blindness or dexterity impairment, so he knew the issues. Likewise, T1 knew about people who struggled to utilize mice in life or from media but had not received formal accessibility education or "learned things seriously." He noted, "My understanding of accessibility is probably not great. I know very little for the most part. It's an area of expertise. I know a very savvy accessibility guy. There's a lot more than I know. I can understand the games, but I'm not an expert."

The teachers all expressed praise for our accessibility education. In addition to the directness expressed by T3, he further commented on how the simulation games added to the limited set of tools for teaching accessibility, "I don't teach other disabilities than color blindness, but I ought to, potentially using your tools. I haven't seen many tools like that." T4 found younger students had fun with the games during the teaching activities. T1 tended to "give students a starter item such as videos and games" before teaching AI. He thought the games we used could be a starter for accessibility education. T2 commented, "It's a great teaching tool. This is perfect. It teaches design thinking as well. Design, CS, and other courses can use it, like building stuff and accommodating disability needs."

## DISCUSSION

Educating next-generation software designers/developers about accessibility is a viable and profound way to instill computational thinking, design thinking, and responsible design/development principles (El-Glaly et al., 2020). This way, more accessible technologies will be designed when accessibility-literate designers enter the IT workforce. Accessibility education has been extensively studied in recent years, but mostly emphasized university education. It was shown to be effective, at least in terms of short-term learning effect (Ludi et al., 2018).

Limited prior literature has addressed accessibility education in high school. To our knowledge, there was only one online module (lecture and quizzes) for teaching high school students about accessibility (Kelly & El-Glaly, 2021). Through a landscape survey of PreK-12 CS Teachers, Blaser et al. (2024) found that 20% of K-12 teachers reported teaching accessibility, yet "accessibility was the least taught computing concept." Toward narrowing this research gap, we integrated research advancement, i.e., simulation games (Kletenik & Adler, 2022, 2023, 2024) into teaching materials and taught accessibility to high school students, in the hope that they are equipped with knowledge and awareness of accessibility and empathy for disabilities and that they can use the knowledge learned for accessible design practices one day.

Students' game-playing data and survey responses revealed similar learning effects compared to CS and non-CS undergraduate students (Kletenik & Adler, 2022). First, simulating disabilities has the potential to help high school students understand the challenges faced by people with disabilities, as evidenced by their negative emotions and lower winning percentages in Round 2 of the games. With accessibility features, they could perform better in disability modes, which may motivate them to explicitly consider accessibility in future design practices. Second, our analysis indicated students' improved empathy for people with disabilities and increased intention to design for them, as indicated by Likert-scale question and open-ended question responses. Third, like college students, high school students felt it was more important to include people with disabilities to vote than game playing. We therefore argue that barriers to the use of technology for people with disabilities are an equity issue and can be framed as such even without detailed technical examples such as WCAG guidelines or technical assignments (Kuang et al., 2024). The similar learning effects suggest the possibility of utilizing games, as well as other pedagogical methods such as experiential learning (El-Glaly et al., 2020) in accessibility education in high school.

Prior research found K-12 CS teachers were willing to teach accessibility (Adler & Kletenik, 2023). Through classroom discussions and teacher interviews, we found that accessibility is not a core part of high school CS curricula. It was taught in elective Web Dev courses but appeared less often in mandatory courses such as Computer Literacy 1 & 2. Accessibility is a social/ethical topic but is not typically seen that way by K-12 teachers. Challenges such as teachers' knowledge of accessibility, conflicted learning goals, and a lack of educational materials exist. Game-based teaching has the potential to narrow this pedagogical gap. Both students and teachers in our study were receptive to the games as educational materials and liked the way of learning accessibility through simulation. Our findings also highlight the need to strengthen high school CS teachers' knowledge of accessibility.

There are several limitations of our study as well as future research directions. First, our education focuses on computer software and website accessibility. Future research may consider developing mobile-compatible games to teach about mobile accessibility. Second, we only conducted the education outreach in a selective admission midwestern public high school, where students were exposed to a wide range of computing concepts. As some teachers suggested, more barriers to teaching accessibility may exist in low-resourced schools. Future research can validate our findings in other types of high schools such as less competitive public schools, private schools, or schools in rural areas. Third, future research might want to understand how students' demographic variables (e.g., gender, race), cultural backgrounds, and prior computer literacy impact learning effect, which is left out of the scope of the current study. Fourth, researchers have shown that educational interventions often failed to incur long-term changes in student attitudes towards accessibility (Zhao et al., 2020). A longitudinal study is helpful to examine the long-term effect of games in accessibility education. Finally, the researchers' presence during the activities might have primed students to give answers that suggested they cared about accessibility. Future research can ask teachers to facilitate the activities without researchers' presence for a comparison.

## CONCLUSION

Through teaching high school students about accessibility through education games, we showed similar learning effects compared to undergraduate CS/non-CS students. The high school students had more empathy for people with disabilities and more intention to design accessibly after playing the games. Interviews with CS teachers in this high school revealed insufficient accessibility education in mandatory CS courses, which is a missed opportunity to equip next-generation software designers/decision-makers with awareness, knowledge, and empathy regarding accessibility. We encourage more research to expand accessibility education in high school by translating accessibility research into educational materials and conducting education outreach activities.

**GENERATIVE AI USE**

We confirm that we did not use generative AI tools/services to author this submission.

**ACKNOWLEDGMENTS**

We would like to thank the high school teachers and students for their participation and the anonymous reviewers for their invaluable feedback. We would like to acknowledge contributions from the following research team members: Bryan Rivera, Elizabeth Lodvikov, and Samantha Sy.

**REFERENCES**

Adler, R. F., & Kletenik, D. (2023). Accessibility for all: Introducing IT accessibility in postsecondary computer science programs for K-12 teachers. *Journal for Postsecondary Education and Disability (JPED)*.

Ali, S., Payne, B. H., Williams, R., Park, H. W., & Breazeal, C. (2019). Constructionism, ethics, and creativity: Developing primary and middle school artificial intelligence education. *International workshop on education in artificial intelligence k-12 (eduai'19)*, *2*, 1–4.

Baker, C., Begel, A., Butler, M., Caspi, A., Ghazal, R., Kingston, N., Lewis, C., Lewis, C., Mack, K., Mbari-Kirika, R., et al. (2023). Accessible computing education in colleges and universities. Retrieved from: https://www.microsoft.com/en-us/research/uploads/prod/2021/02/Accessible-CS-Ed-in-Colleges-and-Universities_Andrew-Begel.pdf

Baker, C. M., El-Glaly, Y. N., & Shinohara, K. (2020). A systematic analysis of accessibility in computing education research. *Proceedings of the 51st ACM Technical Symposium on Computer Science Education*, 107–113.

Blaser, B., Ladner, R. E., Twarek, B., Stefik, A., & Stabler, H. (2024, May). Accessibility and Disability in PreK-12 CS: Results from a Landscape Survey of Teachers. *Proceedings of the 2024 on RESPECT Annual Conference* (pp. 13-20).

Braun, V., & Clarke, V. (2012). *Thematic analysis*. American Psychological Association.

Carter, J. A., & Fourney, D. W. (2007). Techniques to assist in developing accessibility engineers. *Proceedings of the 9th International ACM SIGACCESS Conference on Computers and Accessibility*, 123–130.

Chapman, P., Burket, J., & Brumley, D. (2014). Picoctf: A game-based computer security competition for high school students. *2014 USENIX Summit on Gaming, Games, and Gamification in Security Education (3GSE 14)*.

Chávez, V. C., & Van Wart, S. (2023). "Accessibility is important to everybody": Unpacking students' understanding about accessibility. *Proceedings of the 17th International Conference of the Learning Sciences-ICLS 2023*, pp. 1474-1477.

Cohen, J. (1988). The effect size. *Statistical power analysis for the behavioral sciences*, 77–83.

Conn, P. (2019). *A systematic analysis of accessibility education within computing disciplines*. Rochester Institute of Technology.

Conn, P., Gotfrid, T., Zhao, Q., Celestine, R., Mande, V., Shinohara, K., Ludi, S., & Huenerfauth, M. (2020). Understanding the motivations of final-year computing undergraduates for considering accessibility. ACM *Transactions on Computing Education (TOCE)*, *20*(2), 1–22.

Coverdale, A., Lewthwaite, S., & Horton, S. (2022). Teaching accessibility as a shared endeavour: Building capacity across academic and workplace contexts. *Proceedings of the 19th International Web for All Conference*, 1–5.

Edwards, A., Wright, P., & Petrie, H. (2006). HCI education: We are failing–why. In *Proceedings of HCI Educators Workshop* (Vol. 2006).

El-Glaly, Y., Shi, W., Malachowsky, S., Yu, Q., & Krutz, D. E. (2020). Presenting and evaluating the impact of experiential learning in computing accessibility education. *Proceedings of the ACM/IEEE 42nd International Conference on Software Engineering: Software Engineering Education and Training*, 49–60.

Forsyth, S., Dalton, B., Foster, E. H., Walsh, B., Smilack, J., & Yeh, T. (2021). Imagine a more ethical AI: Using stories to develop teens' awareness and understanding of artificial intelligence and its societal impacts. *2021 Conference on Research in Equitable and Sustained Participation in Engineering, Computing, and Technology (RESPECT)*, 1–2.

Gellenbeck, E. (2005). Integrating accessibility into the computer science curriculum. *Journal of Computing Sciences in Colleges*, *21*(1), 267–273.

Kang, J., DC Chan, A., MJ Trudel, C., Vukovic, B., & Girouard, A. (2021). Diversifying accessibility education: Presenting and evaluating an interdisciplinary accessibility training program. *Proceedings of the 21st Koli Calling International Conference on Computing Education Research*, 1–6.

Kelly, B., & El-Glaly, Y. (2021). Introducing accessibility to high school students. *Proceedings of the 52nd ACM Technical Symposium on Computer Science Education*, 163–169.

Kletenik, D., & Adler, R. F. (2022). Let's play: Increasing accessibility awareness and empathy through games. *Proceedings of the 53rd ACM Technical Symposium on Computer Science Education-Volume 1*, 182–188.

Kletenik, D., & Adler, R. F. (2023). Who wins? A comparison of accessibility simulation games vs. classroom modules. *Proceedings of the 54th ACM Technical Symposium on Computer Science Education*.

Kletenik, D., & Adler, R. F. (2024). Motivated by inclusion: Understanding students' empathy and motivation to design accessibly across a spectrum of disabilities. *Proceedings of the 55th ACM Technical Symposium on Computer Science Education*.

Krippendorff, K. (2018). *Content analysis: An introduction to its methodology*. Sage publications.

Kuang, E., Bellscheidt, S., Pham, D., Shinohara, K., Baker, C. M., & Elglaly, Y. N. (2024). Mapping accessibility assignments into core computer science topics: An empirical study with interviews and surveys of instructors and students. *Proceedings of the CHI conference on Human Factors in Computing Systems*.

Kurniawan, S. H., Arteaga, S., & Manduchi, R. (2010). A general education course on universal access, disability, technology and society. *Proceedings of the 12th international ACM SIGACCESS conference on Computers and accessibility*, 11–18.

Liffick, B. W. (2004). An assistive technology project for an HCI course. *ACM SIGCSE Bulletin*, *36*(3), 273–273.

Lorgat, M. G., Paredes, H., & Rocha, T. (2022). An approach to teach accessibility with gamification. *Proceedings of the 19th International Web for All Conference*, 1–3.

Ludi, S., Huenerfauth, M., Hanson, V., Rajendra Palan, N., & Garcia, P. (2018). Teaching inclusive thinking to undergraduate students in computing programs. *Proceedings of the 49th ACM Technical Symposium on Computer Science Education*, 717–722.

McDonald, N., Schoenebeck, S., & Forte, A. (2019). Reliability and inter-rater reliability in qualitative research: Norms and guidelines for CSCW and HCI practice. *Proceedings of the ACM on human-computer interaction*, 3(CSCW), 1–23.

Nishchyk, A., & Chen, W. (2018). Integrating universal design and accessibility into computer science curricula - A review of literature and practices in Europe. *Studies in health technology and informatics*, *256* (0130), 56–66.

Patricia, L. (2011, May). Service learning: an HCI experiment. In *Proceedings of the 16th Western Canadian Conference on Computing Education* (pp. 12-16).

Putnam, C., Dahman, M., Rose, E., Cheng, J., & Bradford, G. (2016). Best practices for teaching accessibility in university classrooms: Cultivating awareness, understanding, and appreciation for diverse users. *ACM Transactions on Accessible Computing (TACCESS)*, *8*(4), 13.

Rosmaita, B. J. (2006). Accessibility first! A new approach to web design. *Proceedings of the 37th SIGCSE technical symposium on Computer science education*, 270–274.

Shi, W., Moses, H., Yu, Q., Malachowsky, S., & Krutz, D. E. (2023). All: Supporting experiential accessibility education and inclusive software development. *ACM Transactions on Software Engineering and Methodology*, *33*(2), 1–30.

Shinohara, K., Kawas, S., Ko, A. J., & Ladner, R. E. (2018). Who teaches accessibility? A survey of US computing faculty. *Proceedings of the 49th ACM Technical Symposium on Computer Science Education*, 197–202.

Siegfried, R. M., & Leune, K. (2022). A teaching case: A seminar course in accessible computing. *Proceedings of the EDSIG Conference ISSN*, 2473, 4901.

Wald, M. (2008). Design of a 10 credit Masters level assistive technologies and universal design module. *International Conference on Computers for Handicapped Persons*, 190–193.

Zhao, Q., Mande, V., Conn, P., Al-khazraji, S., Shinohara, K., Ludi, S., & Huenerfauth, M. (2020, October). Comparison of methods for teaching accessibility in university computing courses. In *Proceedings of the 22nd International ACM SIGACCESS Conference on Computers and Accessibility* (pp. 1-12).